\documentclass[12pt]{article}
\usepackage{amssymb}
\begin{document}
\hspace*{\fill} LMU--TPW 96--23\\
\hspace*{\fill} hep-th/9609190 \\[2ex]
\section*{}
\begin{center}
\Large\bf 
The dilaton as a candidate for dark matter
\end{center}
\section*{}
\normalsize \rm
\begin{center}
{\bf Rainer Dick}\\[1ex] 
{\small \it Sektion Physik der Universit\"at M\"unchen\\
Theresienstr.\ 37, 80333 M\"unchen, Germany}
\end{center}
\section*{}
{\bf Abstract:} We examine consequences of the stabilization of the
dilaton through the axion.
An estimate of the resulting dilaton
potential yields a relation between
the axion parameter $m_a f_{PQ}$ and the average instanton 
radius, and
predicts the ratio between the dilaton mass $m_\phi$ and the 
axion mass $m_a$.
If we identify the string axion with a Peccei--Quinn axion, 
then $m_\phi m_{Pl} \sim m_a f_{PQ}$, and the dilaton should be strongly
aligned $\sqrt{\langle\phi^2\rangle}\leq 10^{-4}m_{Pl}$ at the QCD scale,
in order not to overclose the universe.

\vspace*{\fill}
\noindent
{\footnotesize Contribution to the Workshop on Aspects 
of Dark Matter in Astro-- and Particle Physics,
Heidelberg (Germany) 16--20 September 1996.\\
email: Rainer.Dick@physik.uni--muenchen.de.}
\newpage
There exists strong experimental and theoretical evidence
for the existence of dark matter in the universe from galactic
rotation curves, gravitational lensing, and standard inflationary
models.
An exciting possibility
for theoreticians and experimentalists alike 
is the existence of non--baryonic dark matter.
Investigation of properties of non--baryonic dark matter and its
cosmological implications is a very interesting 
topic on the forefront both
of astrophysics and particle physics, and concentrates mainly
on massive neutrinos, neutralinos as harbingers of supersymmetry,
and axions, which are predicted from the Peccei--Quinn mechanism to
solve the strong CP problem in QCD \cite{PQ,sw,fw}
and appear also in string theory \cite{GSW}. The books of 
Kolb and Turner
and of B\"orner
provide very useful reviews and
critical discussions of dark matter in cosmology \cite{KT,gb}.

The axion is one of the leading candidates for cold dark matter 
in the universe (besides the lightest supersymmetric particle) 
and very likely
made an important contribution to structure formation in the 
universe\footnote{See \cite{ggr} for a recent review.}.
However, string theory and any theory involving a 
Kaluza--Klein picture
of physics beyond the standard model also predict a dilaton, 
with a characteristic
coupling to gauge fields of the form $\exp(\lambda\phi)F^2$. 
Here $\phi$ denotes
the dilaton and $\lambda$ is a characteristic length, 
typically of the order of
the Planck length\footnote{Our conventions for Planck units are
$m_{Pl}=(8\pi G)^{-1/2}=2.4\times10^{18}$GeV.}, 
which governs all non--gravitational interactions of the dilaton.
Indeed, recent results on strong--weak coupling dualities in 
gauge theories and
string theory provide strong evidence that the axion and the 
dilaton should come
together, and that their properties are intimately connected. From 
these features, it is very likely that the dilaton may 
also serve as a suitable candidate
for cold dark matter, not as a competitor but as a 
companion of the axion. Unfortunately,
discussions of cosmological implications
of a dilaton suffer from the problem of dilaton stability.

String inspired effective field theories and 
Kaluza--Klein theories have an invariance
of the equations of motion under constant 
shifts of the dilaton $\phi\to\phi+c$ and
appropriate rescalings of the other fields. 
Even worse, if the field theory is supersymmetric
the non--renormalization theorem tells us that no 
dilaton mass can be generated 
perturbatively \cite{DS}. This is worrisome, since 
shifts of the dilaton correspond to
rescalings of couplings and masses in the field theory 
and question the meaning
of these parameters. This problem is known as the dilaton problem.

Investigations in the subject of dilaton stabilization
concentrate on the construction of appropriate super--potentials 
or K\"ahler--potentials which stabilize $\langle\exp(\lambda\Phi)\rangle$
at phenomenologically interesting
expectation values, often linking the condensate to a gaugino
condensate and supersymmetry breaking. \cite{gaugino} contains
a list of pioneering papers on this subject. Duality invariant gaugino
condensation and S--duality invariant super--potentials were 
investigated
in \cite{FILQ1,FILQ2}, and
recent contributions can be found in \cite{LNN}, where the coupling 
of the chiral
dilaton multiplet is re--examined, and in \cite{BGW}, where the dilaton
is treated in the linear multiplet.

In the present paper I will point out that topologically 
non--trivial configurations
of the axion and instantons generate
a dilaton potential and consider some consequences of this observation
\cite{mpla3}. However, for the time being I will not examine
what kind of super--potential this mechanism might 
generate in four--dimensional
supersymmetric field theories, but concentrate on the dilaton 
potential and
its cosmological implications.

The main players in the game are a dilaton $\phi$, an axion $a$
and gauge fields $A_\mu$ with field strengths $F_{\mu\nu}$, 
and their mutual interactions before taking into account 
non--perturbative
effects are governed by a Lagrangian
\begin{equation}\label{treeL}
\frac{1}{\sqrt{-g}}{\cal L}=\frac{1}{2\kappa}R-\frac{1}{2}g^{\mu\nu}
\partial_\mu\phi\cdot\partial_\nu\phi
-\frac{1}{2}\exp(-2\lambda\phi)g^{\mu\nu}
\partial_\mu a\cdot\partial_\nu a
\end{equation}
\[
-\frac{1}{4}\exp(\lambda\phi)
F_{\mu\nu}^j F^{\mu\nu}_j +
\frac{q^2}{64\pi^2f_{PQ}}\epsilon^{\mu\nu\rho\sigma} a F_{\mu\nu}^j
F_{\rho\sigma j}
\]

In writing down this Lagrangian, we explicitly split the dilaton
into its expectation value $\langle\exp(\lambda\Phi)\rangle\equiv q^{-2}$
and the fluctuations $\phi$ around the expectation value. We also 
rescaled the gauge fields and the axion such that the 
covariant derivatives 
are $D_\mu =\partial_\mu -igA_\mu$ and the kinetic term for the axion
has standard normalization.

The dilaton--axion and dilaton--gluon couplings in (\ref{treeL})
match in such a way that the system
exhibits an SL(2,$\mathbb R$) strong--weak coupling 
duality in the abelian case \cite{STW}.
S--duality in string theory \cite{FILQ2,SS,sen,js} indicates, 
that the low energy regime shoud also
be described by a field theory with dilaton couplings 
related in this way, and it is a remarkable
property of Kaluza--Klein theory that 
compactifications from $D$ to $d$ dimensions
yield exactly this  S--dual dilaton coupling 
if and only if\footnote{It is
tempting to conclude from this property, that preservation 
of S--duality in the low energy
regime picks out four dimensions.}  $d=4$ \cite{dihab}.

In the case of an abelian gauge group
S--duality fixes the dilaton scale according to
\[
\frac{1}{\lambda}= \frac{8\pi^2}{q^2}f_{PQ}
\]
The resulting SL(2,$\mathbb{R}$) invariance of the equations
of motion is most conveniently described in terms
of the axidilaton
\[
z=\frac{1}{f_{PQ}}[a+\frac{i}{\lambda}\exp(\lambda\phi)]
\]
The symmetry is realized via
\[
z'=\frac{a_{11}z+a_{12}}{a_{21}z+a_{22}},
\qquad\qquad a_{11}a_{22}-a_{12}a_{21}=1
\]
\[
F'_{\mu\nu}-i{\tilde{F}'}_{\mu\nu}
=(a_{21}z+a_{22})(F_{\mu\nu}-i\tilde{F}_{\mu\nu})
\]
which means that the self--dual part of the Yang--Mills curvature
transforms like a half--differential on the axidilaton upper half--plane.

This duality symmetry does not survive in the nonabelian case, 
since the duality
rotation replaces the Bianchi 
identity $D^\mu \tilde{F}_{\mu\nu}=0$ by $D^\mu {\tilde{F}'}_{\mu\nu}=0$,
and it is not possible to identify the rotated field strengths with duality
rotated gauge potentials. 

Perturbatively the nonabelian theory
still has a Peccei--Quinn symmetry $z\to z+a_{12}$
with a conserved current
\[
\frac{1}{\sqrt{-g}}j^\mu=\exp(-2\lambda\phi)g^{\mu\nu}
\partial_\nu a+\frac{q^2}{16\pi^2 f_{PQ}}\epsilon^{\mu\nu\rho\sigma}
(A_\nu^j\partial_\rho A_{\sigma j}+\frac{q}{3}f_{ijk}A_\nu^i A_\rho^j 
A_\sigma^k) 
\]
and the scaling symmetry of the equations of motion, which serves as a
harbinger of the dilaton problem, is also preserved
if the metric is rescaled:
\begin{equation}\label{scale2}
\phi\to\phi+c
\end{equation}
\[
g^{\mu\nu}\to\exp(-\lambda c)g^{\mu\nu}
\]
\[
A_\mu\to A_\mu 
\]
\[
a\to\exp(\lambda c)a
\]

However, we have not yet taken into account 
non--trivial field configurations
of the gauge fields and the axion:
We infer the non--perturbative effects of these field configurations 
from the Lagrangian of the
Euclidean action. In a flat background this takes the form:
\begin{equation}\label{treeLE}
{\cal L}_E=\frac{1}{2}g^{\mu\nu}
\partial_\mu\phi\cdot\partial_\nu\phi
+\frac{1}{2}\exp(-2\lambda\phi)g^{\mu\nu}
\partial_\mu a\cdot\partial_\nu a
\end{equation}
\[
+\frac{1}{4}\exp(\lambda\phi)
F_{\mu\nu}^j F^{\mu\nu}_j +i
\frac{q^2}{32\pi^2f_{PQ}}a \tilde{F}_{\mu\nu}^j
F^{\mu\nu}_j
\]
Positivity of the real part and the estimate of the effective axion
potential by Vafa and Witten \cite{VW}
indicate that the dominating contributions to the path integral come
from instanton configurations $F=\pm\tilde{F}$ with constant dilaton
and the axion frozen to integer multiples of $2\pi f_{PQ}$.
This survival of instantons in the presence of the dilaton is crucial, since
integrality of the instanton number and invariance of the path integral
discretize Peccei--Quinn symmetry 
\[
\frac{a_{12}}{2\pi}\in\mathbb{Z},
\]
thereby also breaking
the scale invariance (\ref{scale2}). The picture emerging from this
observation shows us that
instantons create an effective axion potential with an enumerable
set of equidistant vacua, thus discretizing Peccei--Quinn
symmetry. Discreteness of the axion vacua and the cosine--like
shape of the axion potential displayed below then
implies breaking of the scaling symmetry and lifts the degeneracy
of the dilaton. 

The impact of instantons on the effective axion potential has been
examined by several authors, and the interpretation of instantons
as real time tunneling configurations between gauge theory vacua 
suggests 
\begin{equation}\label{axipot}
V(a) =m^2_a f^2_{PQ}\!\Big(1-\cos(\frac{a}{f_{PQ}})\Big)
\end{equation}
if the instanton gas
is dilute enough to neglect higher order cosine terms \cite{sc,GPY,jk}. 
While initially this result was inferred from semiclassical calculations
of tunneling amplitudes, the same potential can also be derived in a direct
instanton calculation if the wavelength of the axion is large compared
to the instanton size.

The effective axion potential in turn breaks the scale 
invariance (\ref{scale2})
and indicates that the axidilaton--gluon system also lifts the degeneracy
of the dilaton. 
This is obvious in the gauge sector: Instantons
push the dilaton into the strong gauge coupling regime, since the action
of the instantons decreases with decreasing $\langle\phi\rangle$.
However, non--trivial configurations also arise in the axion sector:\\
-- If $a$ is periodic $a\sim a+2\pi f_{PQ}$,
then it contributes local minima to the Euclidean
path integral over $\exp(-S_E)$ in the form of {\sl axion walls}
(instead of axion strings in three dimensions).
Periodicity of $a$ arises, if it is related to the argument
of a complex field with frozen modulus
in the low energy regime. This is e.g.\ the case, if $a$ arises
as the phase of  a determinant of local fermion masses.\\
-- If $a$ is not an angular variable, then all the 
possible vacua $\langle a\rangle=2\pi f_{PQ}n$ are distinct
and we expect three--dimensional domain walls separating four--dimensional
domains where $a$ approximates different vacua.

Both of these topological defects mark regions of non--vanishing
gradients $\partial a$ and favor large values of the dilaton
through the dilaton--axion coupling, thus compensating the effect of the
instantons.

  From these observations we may infer an estimate of the dilaton potential,
if we define a characteristic length $\Delta$ of the axion defects:
In the case of an angular axion $\Delta$ would measure the circumference
of the axion walls, while in the second case 
the four--dimensional domain boundaries are 
extended in three dimensions and have an
average
thickness $\Delta$ in the fourth direction.
If $\varrho$ denotes both the
average extension
and separation of instantons we find an estimate
for the effective dilaton potential
\begin{equation}\label{dilpot1}
V(\phi)=\frac{48}{q^2\varrho^4}\exp(\lambda\phi)
+2\pi^2\frac{f_{PQ}^2}{\Delta^2}
\exp(-2\lambda\phi)
\end{equation}
and this implies for the gauge coupling
\begin{equation}\label{consist1}
\langle\exp(\lambda\Phi)\rangle=\frac{1}{q^2}
=\frac{\pi^2 f_{PQ}^2\varrho^4}{12\Delta^2}
\end{equation}

The potential (\ref{dilpot1}) then yields a dilaton mass
\begin{equation}\label{dilmass}
m_\phi=\frac{12\lambda}{q\varrho^2} 
\end{equation}

If $a$ is not an angular
variable we may estimate the parameter $\Delta$ by minimizing
the energy density of the axion domain spaces
\begin{equation}\label{edensds}
u=2\pi^2\frac{f_{PQ}^2}{\Delta}+m_a^2 f_{PQ}^2\Delta.
\end{equation}
This yields a thickness of the order
\begin{equation}\label{dsthick}
\Delta\simeq\frac{\pi\sqrt{2}}{m_a}
\end{equation}
which is of the same order as the thickness of ordinary axion domain walls
in Minkowski space \cite{KT}. From (\ref{consist1}) 
and (\ref{dsthick}) we find a relation between the axion
parameters and the average instanton radius
\begin{equation}\label{rela1}
m_a^2 f_{PQ}^2\simeq\frac{6}{\pi\alpha_q\varrho^4}
\end{equation}

The average instanton radius is set by the QCD scale \cite{GPY}
\[
\frac{1}{\varrho}\sim\Lambda_{\mbox{\footnotesize QCD}}
\sim 2\times 10^8\mbox{eV}
\]
whereas from \cite{BT,KSS} we learn that
\[
m_a f_{PQ}\sim m_{\pi}f_{\pi}\sim 10^{16}\mbox{eV}^2
\]
Relation (\ref{rela1}) is in gross agreement with these
estimates if $\alpha_q\sim O(10)$.

The potential (\ref{dilpot1}) then implies a relation between the 
dilaton mass
and the axion mass
\begin{equation}\label{dilmass2}
m_\phi\simeq\sqrt{6}\lambda m_a f_{PQ}
\end{equation}
which hints at a non--perturbatively generated dilaton mass which is 
much smaller
than the axion mass:
\[
m_\phi\sim 10^{-6}m_a
\]

However, assuming $\lambda\simeq m_{Pl}^{-1}$,
we encounter a fine--tuning problem: The dilaton behaves very similar
to a misalignment produced light axion, and the estimates of the 
cosmic abundance
of the axion \cite{PWW,AS,DF,mst} carry over to the dilaton with 
some minor modifications.
Since the temperature where a misaligned field starts 
to oscillate goes with $T \sim m^{0.18}$
\cite{mst,KT}, the dilaton will start to oscillate after the axion, 
when the temperature
has dropped by another factor of 10. Then we find for the dilaton 
contribution to the
critical density of the universe
\begin{equation}\label{omega}
\frac{\Omega_\phi}{\Omega_a}
\simeq 10^{-5}\frac{\langle\phi^2\rangle}{f_{PQ}^2}\simeq
10^7\frac{\langle\phi^2\rangle}{m_{Pl}^2}
\end{equation}

In the derivation of this ratio we assumed that the dilaton is 
non--relativistic
at the onset of oscillations. This is a justified assumption, 
since the drop in the
temperature relative to the onset of the axion oscillations 
implies for the ratio
of the momenta $p_\phi\simeq 10^{-2}p_a$, yielding
a velocity $p_\phi/m_\phi\simeq 10^{-2}$.

Imposing a 
bound $\Omega_\phi/\Omega_a \leq 0.1$ then implies fairly strong
alignment of the dilaton: 
\[
\sqrt{\langle\phi^2\rangle}\leq 10^{-4}m_{Pl}
\]
Eventually, this could be attributed to inflation of a local 
patch with a small variance
of the dilaton, which would also explain the absence of 
variations in spectral lines
over cosmic distances:
Patches with small variation of the dilaton survive and inflate, 
while overclosed patches
collapse. The variation of the dilaton in the inflating
patch remains small, because the dilaton
is very weakly coupled.

We have encountered a generic problem of Planck scale physics:
Light moduli either must have a decay constant far below the 
Planck scale,
or they must be strongly aligned when oscillation 
begins to dominate, 
in order
not to overclose the universe. Recent proposals for 
accomodating 
very small decay constants in terms of Planck units 
are discussed in
\cite{BD} (see also \cite{CK}), and eventually we 
may also find a mechanism to
considerably lower the decay constant of the dilaton. 
This would not alter
our conclusions about the mechanism of 
dilaton stabilization, 
but could imply a more
conventional picture of the cosmological 
significance of the dilaton.


\begin{thebibliography}{88}
 \bibitem[1]{PQ} R.D.\ Peccei and H.R.\ Quinn, {\sl Phys.\ Rev.\ Lett.\ }{\bf 38}
 (1977) 1440; {\sl Phys.\ Rev.\ }{\bf D16} (1977) 1791.
 \bibitem[2]{sw} S.\ Weinberg, {\sl Phys.\ Rev.\ Lett.\ }{\bf 40} (1978) 223.
 \bibitem[3]{fw} F.\ Wilczek, {\sl Phys.\ Rev.\ Lett.\ }{\bf 40} (1978) 279.
 \bibitem[4]{GSW} M.B.\ Green, J.H.\ Schwarz and E.\ Witten, {\it Superstring Theory},
 2 Vols., Cambridge University Press, Cambridge 1987.
 \bibitem[5]{KT} E.W.\ Kolb and M.S.\ Turner, {\it The Early Universe}, 
 Addison--Wesley,
 Reading 1990.
 \bibitem[6]{gb} G.\ B\"orner, {\it The Early Universe}, $3^{\mbox{\footnotesize rd}}$ ed.,
 Springer, Berlin 1993.
 \bibitem[7]{ggr} G.G.\ Raffelt, {\it Axions in astrophysics and cosmology},
 hep--ph/9502358.
 \bibitem[8]{DS} M.\ Dine and N.\ Seiberg, 
 {\sl Phys.\ Rev.\ Lett.\ }{\bf 57} (1986) 2625.
 \bibitem[9]{gaugino} G.\ Veneziano and S.\ Yankielowicz, {\sl Phys.\ Lett.\ }{\bf B113}
 (1982) 231;\\
 S.\ Ferrara, L.\ Girardello and H.P.\ Nilles, {\sl Phys.\ Lett.\ }{\bf B125} (1983) 457;\\
 J.P.\ Derendinger, L.E.\ Iba\~{n}ez and H.P.\ Nilles, {\sl Phys.\ Lett.\ }{\bf B155}
 (1985) 467;\\
 M.\ Dine, R.\ Rohm, N.\ Seiberg and E.\ Witten, {\sl Phys.\ Lett.\ }{\bf B156}
 (1985) 55;\\
 T.R.\ Taylor, {\sl Phys.\ Lett.\ }{\bf B164} (1985) 43.
 \bibitem[10]{FILQ1} A.\ Font, L.\ Iba\~{n}ez, D.\ L\"ust and F.\ Quevedo,
 {\sl Phys.\ Lett.\ }{\bf B 245} (1990) 401.
 \bibitem[11]{FILQ2} A.\ Font, L.\ Iba\~{n}ez, D.\ L\"ust and F.\ Quevedo, 
 {\sl Phys.\ Lett.\ }{\bf B249} (1990) 35.
 \bibitem[12]{LNN} Z.\ Lalak, A.\ Niemeyer and H.P.\ Nilles, 
 {\sl Nucl.\ Phys.\ }{\bf B453}
 (1995) 100.
 \bibitem[13]{BGW} P.\ Bin\'{e}truy, M.K.\ Gaillard and Y.-Y.\ Wu,
 {\it Dilaton stabilization in the context of dynamical supersymmetry breaking
 through gaugino condensation}, hep--th/9605170.
 \bibitem[14]{mpla3} R.\ Dick, {\it Stabilizing the dilaton through the axion},
 LMU--TPW 96--22, submitted to {\sl Mod.\ Phys.\ Lett.\ }{\bf A}.
 \bibitem[15]{STW} A.\ Shapere, S.\ Trivedi and F.\ Wilczek,  
 {\sl Mod.\ Phys.\ Lett.\ }{\bf A6} (1991) 2677.
 \bibitem[16]{SS} J.H.\ Schwarz and A.\ Sen, {\sl Nucl.\ Phys.\ }{\bf B411} (1994) 35;
 {\sl Phys.\ Lett.\ }{\bf B312} (1993) 105.
 \bibitem[17]{sen} A.\ Sen, {\sl Int.\ J.\ Mod.\ Phys.\ }{\bf A9} (1994) 3707; 
 {\sl Phys.\ Lett.\ }{\bf B329} (1994) 217.
 \bibitem[18]{js} J.H.\ Schwarz, {\it Lectures on superstring and M theory dualities},
 hep--th/9607201.
 \bibitem[19]{dihab} R.\ Dick, 
 {\it The dilaton in gravity and cosmology}, in preparation.
 \bibitem[20]{VW} C.\ Vafa and E.\ Witten, {\sl Phys.\ Rev.\ Lett.\ }{\bf 53} (1984)
 535.
 \bibitem[21]{sc} S.\ Coleman, in {\it The Whys of Subnuclear Physics}, A.\ Zichichi (Ed.),
 Plenum Press, New York 1979, pp.\ 805--916.
 \bibitem[22]{GPY} D.J.\ Gross, R.D.\ Pisarski and L.G.\ Yaffe, 
 {\sl Rev.\ Mod.\ Phys.\ }{\bf 53} (1981)
 43.
 \bibitem[23]{jk} J.E.\ Kim, {\sl Phys.\ Rep.\ }{\bf 150} (1987) 1.
 \bibitem[24]{BT} W.A.\ Bardeen and S.-H.H.\ Tye, {\sl Phys.\ Lett.\ }{\bf B74}
 (1978) 229.
 \bibitem[25]{KSS} J.\ Kandaswamy, P.\ Salomonson and J.\ Schechter,
 {\sl Phys.\ Rev.\ }{\bf D17} (1978) 3051; {\sl Phys.\ Lett.\ }{\bf B74}
 (1978) 377.
 \bibitem[26]{PWW} J.\ Preskill, M.B.\ Wise and F.\ Wilczek, 
 {\sl Phys.\ Lett.\ }{\bf B120} (1983) 127.
 \bibitem[27]{AS} L.F.\ Abbott and P.\ Sikivie, 
 {\sl Phys.\ Lett.\ }{\bf B120} (1983) 133.
 \bibitem[28]{DF} M.\ Dine and W.\ Fischler, 
 {\sl Phys.\ Lett.\ }{\bf B120} (1983) 137.
 \bibitem[29]{mst} M.S.\ Turner, {\sl Phys.\ Rev.\ }{\bf D33} (1986) 889.
 \bibitem[30]{BD} T.\ Banks and M.\ Dine, {\it The cosmology of string theoretic axions},
 hep--th/9608197.
 \bibitem[31]{CK} K.\ Choi and J.E.\ Kim, {\sl Phys.\ Lett.\ }{\bf B176} (1986) 103.

\end{thebibliography}
\end{document}